\documentclass[conference]{IEEEtran}
\IEEEoverridecommandlockouts

\usepackage{cite}
\usepackage{amsmath,amssymb,amsfonts}
\usepackage{algorithmic}
\usepackage{graphicx}
\usepackage{textcomp}
\usepackage{xcolor}
\usepackage{subfigure}
\usepackage{subfig}
\usepackage{multirow}
\usepackage{supertabular,booktabs}
\usepackage{colortbl}
\usepackage{soul} 

\def\BibTeX{{\rm B\kern-.05em{\sc i\kern-.025em b}\kern-.08em
    T\kern-.1667em\lower.7ex\hbox{E}\kern-.125emX}}
\makeatletter
\newcommand{\linebreakand}{%
  \end{@IEEEauthorhalign}
  \hfill\mbox{}\par
  \mbox{}\hfill\begin{@IEEEauthorhalign}
}
\makeatother

\begin{document}
\include{bibliography}

\title{Assessing the Reliability and Validity of a Balance Mat for Measuring Postural Stability: A Combined Robot-Human Approach
}

\author{
    \IEEEauthorblockN{
        Abishek Shrestha\IEEEauthorrefmark{1}, 
        Damith Herath\IEEEauthorrefmark{1}, 
        Angie Fearon\IEEEauthorrefmark{2}, 
        Maryam Ghahramani\IEEEauthorrefmark{1}
    }
    \IEEEauthorblockA{\IEEEauthorrefmark{1}Faculty of Science and Technology, University of Canberra, Australia}
    \IEEEauthorblockA{\IEEEauthorrefmark{2}Faculty of Health, University of Canberra, Australia}
    \IEEEauthorblockA{Emails: \{Abishek.Shrestha, Damith.Herath, Angie.Fearon, Maryam.Ghahramani\}@canberra.edu.au}
    \vspace{-3em}
}

\maketitle
\footnotetext{\textcopyright 2025 IEEE. Accepted for publication in IEEE Medical Measurements And Applications (MeMeA). Personal use of this material is permitted. Permission from IEEE must be obtained for all other uses.
}
\begin{abstract}
Postural sway assessment is important for detecting balance problems and identifying people at risk of falls. Force plates (FP) are considered the gold standard postural sway assessment method in laboratory conditions, but their lack of portability and requirement of high-level expertise limit their widespread usage. This study evaluates the reliability and validity of a novel Balance Mat (BM) device, a low-cost portable alternative that uses optical fibre technology. The research includes two studies: a robot study and a human study. In the robot study, a UR10 robotic arm was used to obtain controlled sway patterns to assess the reliability and sensitivity of the BM. In the human study, 51 healthy young participants performed balance tasks on the BM in combination with an FP to evaluate the BM's validity. Sway metrics such as sway mean, sway absolute mean, sway root mean square (RMS), sway path, sway range, and sway velocity were calculated from both BM and FP and compared. Reliability was evaluated using the intra-class correlation coefficient (ICC), where values greater than 0.9 were considered excellent and values between 0.75 and 0.9 were considered good. Results from the robot study demonstrated good to excellent ICC values in both single and double-leg stances. The human study showed moderate to strong correlations for sway path and range. Using Bland-Altman plots for agreement analysis revealed proportional bias between the BM and the FP where the BM overestimated sway metrics compared to the FP. Calibration was used to improve the agreement between the devices. The device demonstrated consistent sway measurement across varied stance conditions, establishing both reliability and validity following appropriate calibration. 
\end{abstract}


\section{Introduction}

Postural sway is the subtle change in posture that occurs while standing still, and it serves as an important indicator of balance control and fall risk \cite{noohu2014relevance}. Falls and associated injuries from resulting trauma present a serious public health issue. For those over 65 years old, falls are a leading cause of injury-related hospitalisations, increased healthcare needs, and mortality  \cite{levy2018validity}. Postural sway assessment is important for identifying balance problems at early stages and developing rehabilitation plans to tackle those issues \cite{guskiewicz1996research, noohu2014relevance}. The traditional clinical balance tests like the Berg Balance Scale \cite{berg1992measuring}, Tinetti Performance Oriented Mobility Assessment \cite{tinetti1986performance}, Timed up and go \cite{podsiadlo1991timed} are widely used but are limited by their inability to provide quantitative motion data \cite{levy2018validity}. Force plates (FP) are considered the gold standard postural sway assessment method in laboratory conditions, although they lack portability, require expertise to interpret the data and require an external power source\cite{richmond2018leveling}. This has led to a move towards portable and low-cost instrumented tools to measure balance in clinical and home settings.\\
In ensuring that the balance assessment devices produce clinically relevant results the reliability and validity of the devices should be determined. Reliability assesses the extent to which measurements are consistent, while validity checks how well the device measures what it has been designed to measure\cite{kimberlin2008validity}. The Balance Mat (BM)(Balance Mat Pty Ltd, Canberra, Australia), a portable device designed for postural sway assessment, uses optical fibre technology underneath the mat to quantify pressure distribution exerted by the foot in the form of voltage signals during various stance conditions. It is necessary to clarify the performance of BM both in terms of reliability and validity to determine if it is suitable for use in a clinical setting as a tool for evaluating balance and measuring postural sway\cite{Ghahramani2022}. 

This study seeks to evaluate the BM through two key objectives: (1) evaluating its test-retest reliability by using a robotic arm to simulate controlled sway patterns, and (2) examining its validity by comparing postural sway measurements obtained from 51 healthy young participants to the sway data obtained from the FP on different stance conditions. 

The first research contribution of this study is the demonstration that BM can produce reliable postural sway measurements in multiple trials, as shown by good to excellent intraclass correlation coefficients (ICC) in the Robot study. Furthermore, by comparing its performance with the FP, the BM demonstrated moderate to high correlations in various stance conditions in the human study. These results align with those of Ghahramani et al. \cite{Ghahramani2022} that the BM offers reliable sway measurements and also provides a valid, low-cost, and portable solution for balance assessment, making it an accessible tool for clinical settings. 
 
\section{Study Design} \subsection{Instrumentation} The BM is an Australian-made, patented balance testing device approved by the Therapeutic Goods Administration (TGA). It uses plastic optical fibre arranged in a grid, with 32 to 40 crossover points acting as individual sensors under the feet. When pressure is applied, the compression of the fibre alters the intensity of light transmitted through it. The microcontroller at one end of the mat detects this change in light, which is converted into a pressure signal. These signals are transmitted to a computer via USB and do not require an external power source. The BM records real-time postural sway data at a sampling frequency of 40 Hz and is unitless. It is a portable device, measuring 600mm x 700mm x 6mm and weighing 2.5 kg\cite{Ghahramani2022}.

\begin{figure}
    \centering
    \includegraphics[width=\textwidth,height=0.4\textwidth,keepaspectratio]{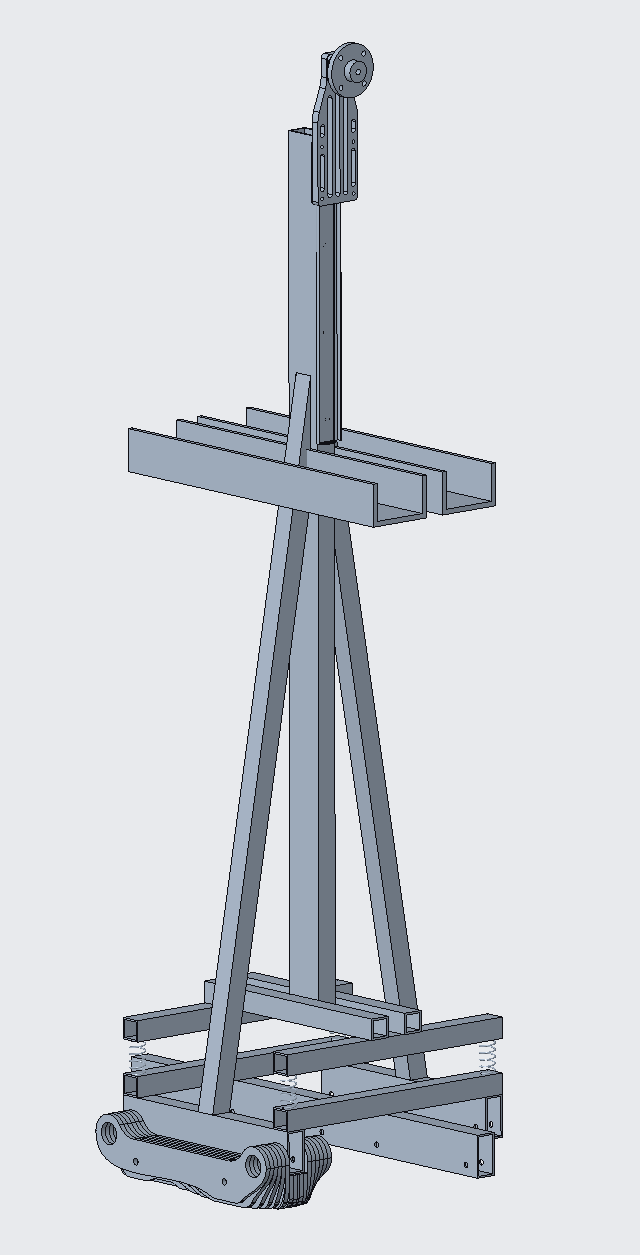}
    \caption{Computer Aided Design (CAD) of the test rig used in the robot study}
    \label{fig:rig}
    \vspace{-1.1em}
\end{figure}

For the robot study, a Universal Robots UR10 robotic arm (Universal Robots, Odense, Denmark) was employed to simulate controlled and repeatable sway patterns. The robotic arm was programmed in Python to replicate human postural sway, following a defined path where it moved sequentially between designated points (Figure \ref{fig: VisualizationRobotSwayPath}). A custom-built rig, as shown in Figure~\ref{fig:rig} was attached to the robotic arm’s end effector, simulating both single leg and double leg stances for the assessment of sway under various loading conditions, and placed on top of the BM, which collected raw data at a rate of 40 Hz. This method allowed for high precision and repeatability, key factors for testing the reliability of the BM \cite{leach2014validating}.

The Human Study employed a force plate (Kistler, Model 9260AA6) as the benchmark for balance measurements. The FP captured centre of pressure (CoP) displacements in anterior-posterior (AP) and medial-lateral (ML) directions with high accuracy. The BM was placed on the FP to synchronise data acquisition, with both devices operating at 40 Hz.
\subsection{Study Methods} 
\begin{figure}
    \centering
    \includegraphics[width=0.48\textwidth]{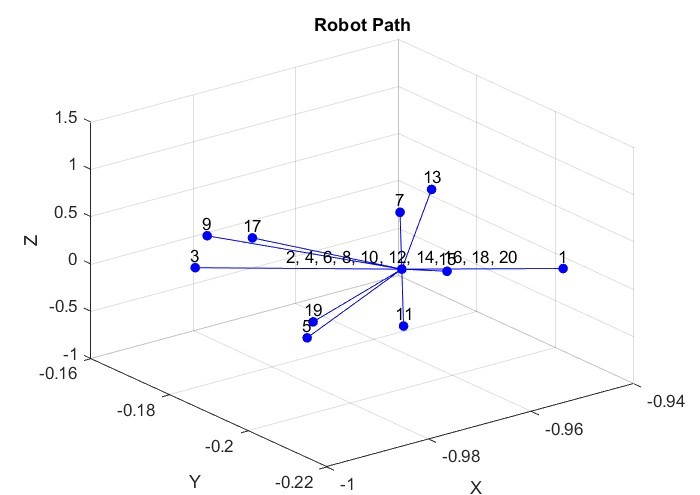}
    \caption{Three-dimensional plot showing the simulated sway path generated by the robotic arm in x, y and z coordinates}
   \label{fig: VisualizationRobotSwayPath}
   \vspace{-1em}
\end{figure}

\begin{figure}[!ht]
  \centering
  \includegraphics[width=\textwidth,height=0.4\textwidth,keepaspectratio]{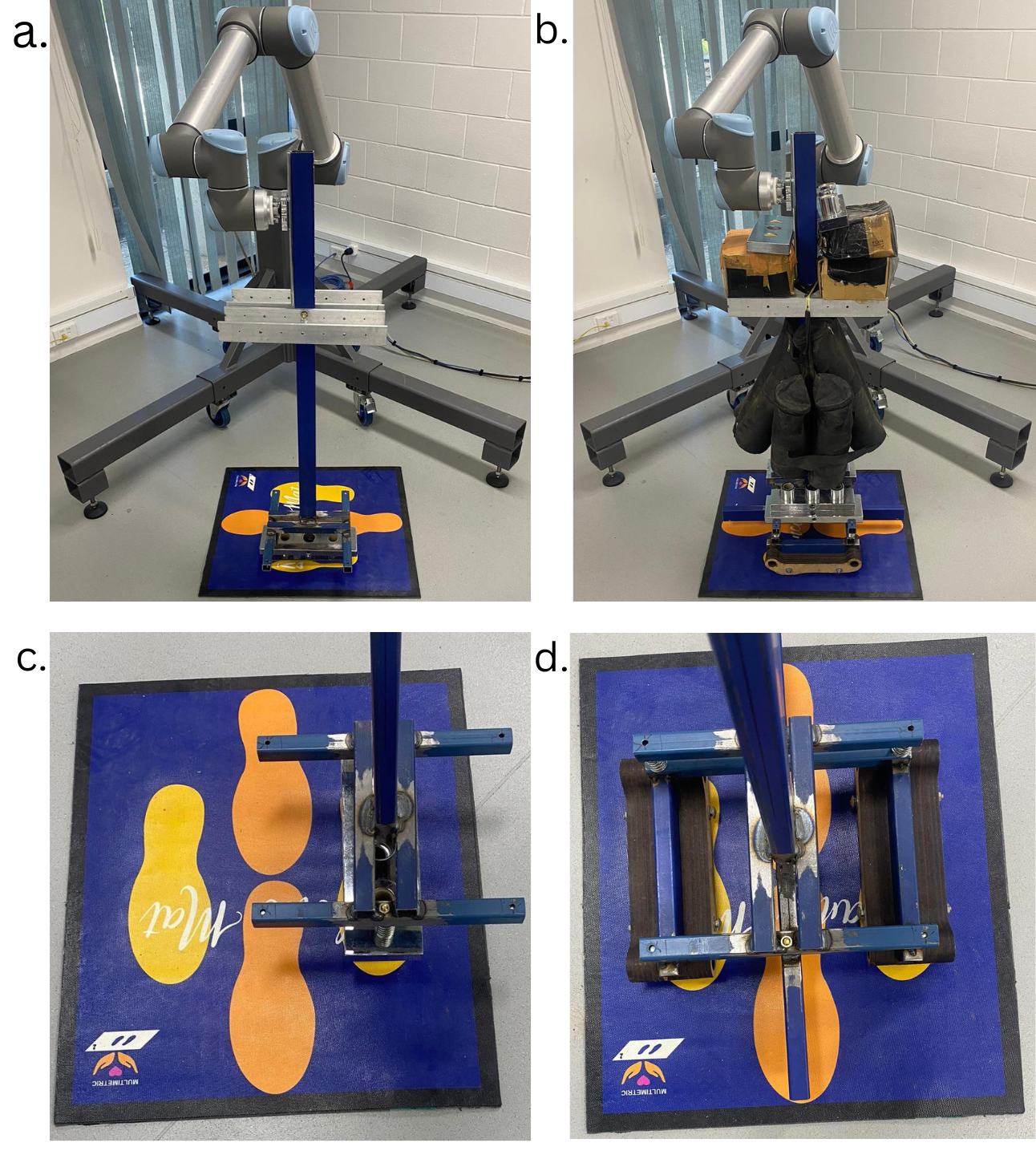}
  \caption{(a) Single leg stance on Robotic Arm (Universal Robots) without loading and (b) Double leg stance with loading condition (c) Single leg stance on BM (d) double leg stance on BM.}
  \vspace{-1.1em}
   \label{fig:RoboticArmSLDLAndLoading}
\end{figure}

\paragraph{Robot Study} The BM’s reliability and sensitivity were assessed using controlled sway patterns generated by the UR10 robotic arm (Figure \ref{fig: VisualizationRobotSwayPath}). Loads ranging from 10 kg to 110 kg, in 10 kg increments, were applied under single-leg and double-leg stance conditions (Figure \ref{fig:RoboticArmSLDLAndLoading}). Fifty 20-second trials were conducted per load and stance configuration, resulting in 1,100 trials. The protocol was repeated for test-retest reliability assessment. Regression modelling analysed the relationship between increased mass and sway parameters.
\paragraph{Human Study} Sample size calculation was done following previous study \cite{fritz2012effect} with a medium effect size (0.18), an alpha level of 0.05, and a statistical power of 0.8, which resulted in a sample size of 46 participants. To account for a dropout rate of 10\%, the final sample size was adjusted to 51 participants. The study included 51 healthy adults (26.8 ± 6 years) with no musculoskeletal or neurological impairments. Exclusion criteria also included lower limb injuries, neuromuscular disorders, and vertigo. Each participant performed a 30-second balance assessment with a 5-second adjustment period under the following stance conditions:
\begin{enumerate} 
    \item Normal stance, feet shoulder-width apart, eyes open 
    \item Normal stance, feet shoulder-width apart, eyes closed 
    \item Tandem stance, eyes open 
    \item Tandem stance, eyes closed 
    \item Single-leg stance (left foot), eyes open 
    \item Single-leg stance (right foot), eyes open 
    \item Double-leg stance, feet together, eyes open 
\end{enumerate}
Simultaneous BM and FP recordings helped in the comparison of sway metrics under different sensory and postural conditions \cite{leach2014validating}. This study was approved by the University of Canberra Human Research Ethics Committee (approval number: 20249208).
\subsection{Study Metrics} Key sway metrics were derived from BM raw data and FP CoP measurements, including sway mean, sway path, sway range, sway root mean square (RMS) and sway velocity \cite{prieto1996measures, paillard2015techniques}. The resultant Vector Distance CoP ($CoP_{RD}$) were calculated using $CoP_{AP}$ and $CoP_{ML}$ for comparative analysis.

\textbf{Metrics Calculation:} \begin{itemize}\setlength\itemsep{0.5em} \item \textbf{$CoP_{RD}$:}\begin{equation} CoP_{RD} = \sqrt{CoP_{AP_i}^2 + CoP_{ML_i}^2} \end{equation}
\item \textbf{Sway Mean:} 
\begin{equation}
Mean\ Sway = \frac{1}{n} \sum_{i=1}^{n} (X_i)
\end{equation}
\item \textbf{Sway RMS:} 
\begin{equation}
RMS\ Sway = \sqrt{\frac{1}{n} \sum_{i=1}^{n} (X_i)^2}  \end{equation}
\item \textbf{Sway Path:} 
\begin{equation}
Sway\ Path = \sum_{i=1}^{n-1} \sqrt{(X_{i+1} - X_i)^2}
\end{equation}

\item \textbf{Sway Range:} 
\begin{equation}
Sway\ Range = \text{Max Sway} - \text{Min Sway}
\end{equation}

\item \textbf{Sway Velocity:} 
\begin{equation}
Sway\ Velocity = \frac{Sway\ Path}{T}
\end{equation}
where $T$ is the trial duration and $X_i$ represents BM raw data or FP CoP values.
\end{itemize}
These metrics facilitated the evaluation of BM’s reliability and validity against the FP, providing insights into postural stability across different stance conditions.

Higher values in these metrics are typically associated with reduced balance control and an increased risk of falling \cite{Mancini2012isway, muir2013dynamic, fernie1982relationship}. Sway path and sway range are found to be useful in discriminating fallers and non-fallers \cite{sturnieks2011validity}.

\subsection{Statistical Analysis}
In the robot study, we used descriptive statistics to examine the sway data across different loading conditions and excluded any abnormal or missing values. ICC with a two-way mixed effects model was used to assess the test-retest reliability of the BM. The coefficient of determination ($R^2$) was used to determine the best fit for various regression models such as linear, cubic, quadratic, and exponential to understand the relationship between increased mass and sway metrics. In the human study, we used the Shapiro-Wilk test to assess normality for both BM and FP sway metrics. Since most parameters were not normally distributed, we used non-parametric statistical methods to evaluate associations. The strength of the correlation coefficient was interpreted as follows: values between 0 and $|$0.09$|$ were considered negligible, $|$0.10$|$ and $|$0.39$|$ weak, $|$0.40$|$ and $|$0.69$|$ moderate, $|$0.70$|$ and $|$0.89$|$ strong, and $|$0.90$|$ and $|$1.00$|$ very strong for predicting relationships \cite{schober2018correlation}. Bland-Altman plots were generated to assess agreement and identify systematic bias or proportional errors. Simple linear regression equations were used to calibrate BM data against FP measurements to improve the agreement of BM with the measures obtained from the FP. All statistical analyses were conducted using IBM SPSS Statistics (IBM Corp.), R Studio (Posit Software, Boston, Massachusetts, United States) and Microsoft Excel (Microsoft Corp., Washington, United States) to provide a comprehensive evaluation of the BM’s reliability and validity for postural sway assessment.

\section{Results}

\subsection{Robot Study}
\subsubsection{Measurement Reliability and Consistency }

Test-retest reliability evaluates the consistency of measurements made on a system across repeated trials under identical conditions and was evaluated using ICC. High reliability indicates minimal variability between results. ICC values approaching 1 signify greater reliability. The test-retest reliability of the device in measuring various postural sway parameters was assessed using ICCs with a two-way mixed effects model and a consistency definition. Reliability was evaluated for mean, absolute mean, RMS, sway path, sway range, sway velocity, and standard deviation. The results demonstrated high reliability across most parameters as shown in Table \ref{tab:ICCTestRetest}.

\begin{table*}[ht]
    \centering
    \caption{Test-Retest Reliability (ICC) for Postural Sway Parameters}
    \label{tab:ICCTestRetest}
    \begin{tabular}{|l|c|c|c|c|c|c|}
    \hline
    \textbf{Parameter} \& \textbf{Stance} \& \textbf{Single Measures ICC (95\% CI)} \& \textbf{Average Measures ICC (95\% CI)} \\
    \hline
    $|$Mean$|$ \& Double \& 0.981** (0.978, 0.984) \& 0.991** (0.989, 0.992) \\
    \hline
    Mean \& Double \& 0.638** (0.585, 0.685) \& 0.779** (0.738, 0.813) \\
    \hline
    RMS \& Double \& 0.987** (0.985, 0.989) \& 0.994** (0.993, 0.995) \\
    \hline
    Sway Path \& Double \& 0.981** (0.978, 0.984) \& 0.991** (0.989, 0.992) \\
    \hline
    Sway Range \& Double \& 0.991** (0.989, 0.992) \& 0.995** (0.994, 0.996) \\
    \hline
    Sway Velocity \& Double \& 0.981** (0.978, 0.984) \& 0.991** (0.989, 0.992) \\
    \hline
    $|$Mean$|$ \& Single \& 0.808** (0.777, 0.835) \& 0.894** (0.875, 0.910) \\
    \hline
    RMS \& Single \& 0.808** (0.777, 0.836) \& 0.894** (0.875, 0.910) \\
    \hline
    Sway Path \& Single \& 0.808** (0.777, 0.835) \& 0.894** (0.875, 0.910) \\
    \hline
    Sway Range \& Single \& 0.845** (0.819, 0.867) \& 0.916** (0.900, 0.929) \\
    \hline
    Sway Velocity \& Single \& 0.808** (0.777, 0.835) \& 0.894** (0.875, 0.910) \\
    \hline
    \multicolumn{4}{l}{\footnotesize{$^{**}$ Statistically significant at $p < 0.001$.}} \\
    \end{tabular}
    \vspace{-1.5em}
    
\end{table*}

\subsubsection{Association Between the Mass and Sway Parameters} 
 As we added weight, the sway mean in the single leg stance ranged from 0.31 (SD = 0.072) for 10 kg to 0.40 (SD = 0.685) for 110 kg in Round 1, and from 0.37 (SD = 0.158) to 0.37 (SD = 1.002) in Round 2. In the double leg stance, sway mean ranged from 0.39 (SD = 0.170) for 10 kg to 0.69 (SD = 0.179) for 110 kg in Round 1, and from 0.4 (SD = 0.035) for 10 kg to 0.69 (SD = 0.146) for 110 kg in Round 2.
 
Regression analysis revealed that cubic models best described the relationship between mass and sway parameters, with $R^2$ values exceeding 0.97 in both single leg and double leg stances, as shown in Table \ref{Table:R2Values} for most of the sway parameters.

\begin{table*}[ht]
    \centering
    \caption{Relationship between Simulated Mass and Postural Sway Parameters ($R^2$ values) in Double Leg and Single Leg Stance Conditions across Two Rounds of Experiments}
    \label{Table:R2Values}
    \begin{tabular}{|l|c|c|c|c|}
    \hline
    \multirow{2}{*}{\textbf{Sway Parameters}} & \multicolumn{2}{c|}{\textbf{Double Leg}} & \multicolumn{2}{c|}{\textbf{Single Leg}} \\ \cline{2-5} 
     \& \textbf{Round 1 ($R^2$)} \& \textbf{Round 2 ($R^2$)} \& \textbf{Round 1 ($R^2$)} \& \textbf{Round 2 ($R^2$)} \\ \hline
    \textbf{$|BM Mean|$ vs Mass} \& 0.979** \& 0.977** \& 0.994** \& 0.997** \\ \hline
    \textbf{BM Mean vs Mass} \& 0.730** \& 0.559** \& 0.357** \& 0.048** \\ \hline
    \textbf{BM Sway RMS vs Mass} \& 0.982** \& 0.978** \& 0.991** \& 0.996** \\ \hline
    \textbf{BM Sway Path vs Mass} \& 0.979** \& 0.977** \& 0.994** \& 0.997** \\ \hline
    \textbf{BM Sway Range vs Mass} \& 0.990** \& 0.987** \& 0.977** \& 0.993** \\ \hline
    \textbf{BM Sway Velocity vs Mass} \& 0.979** \& 0.977** \& 0.994** \& 0.997** \\ \hline
    \multicolumn{4}{l}{\footnotesize{$^{**}$ Statistically significant at $p < 0.001$.}} \\
    \end{tabular}
    \vspace{-1.5em}
\end{table*}

\subsection{Human Study}
\subsubsection{Correlation Analysis}
Tables \ref{tab:FPVSBMEO} and \ref{tab:FPVSBMEC} present the correlation analysis between the BM and the FP under conditions of Eyes Open and Eyes Closed, respectively. 

\begin{table*}[ht]
    \centering
    \caption{Spearman's rank correlation coefficients between BM sway parameters and corresponding CoP measurements from the FP under the Eyes Open condition.}
    \label{tab:FPVSBMEO}
    \begin{tabular}{|l|c|c|c|c|c|c|}
    \hline
    \textbf{FP Parameter} \& \textbf{BM $|$Mean$|$} \& \textbf{BM RMS} \& \textbf{BM Sway Path} \& \textbf{BM Sway Range} \& \textbf{BM Sway Velocity} \\ 
    \hline
    \textbf{$CoP_{AP}$} \& 0.275** \& 0.287** \& 0.730** \& 0.641** \& 0.730** \\ 
    \hline
    \textbf{$CoP_{ML}$} \& 0.051 \& 0.068 \& 0.647** \& 0.657** \& 0.647**\\ 
    \hline
    \textbf{$CoP_{RD}$} \& 0.322** \& 0.322** \& 0.698** \& 0.558** \& 0.698** \\ 
    \hline
    \multicolumn{6}{l}{$^{**}$ Correlation is statistically significant at the 0.001 level (2-tailed) $(p < 0.001)$.}
    \end{tabular}
    \vspace{-1.5em}
    \centering
\end{table*}

\begin{table*}[ht]
    \centering
    \caption{Spearman's rank correlation coefficients between BM sway parameters and corresponding CoP measurements from the FP under the Eyes Closed condition.}
    \label{tab:FPVSBMEC}
    \begin{tabular}{|l|c|c|c|c|c|c|}
    \hline
    \textbf{FP Parameter} \& \textbf{BM $|$Mean$|$} \& \textbf{BM RMS} \& \textbf{BM Sway Path} \& \textbf{BM Sway Range} \& \textbf{BM Sway Velocity}\\ 
    \hline
    \textbf{$CoP_{AP}$} \& 0.370** \& 0.451** \& 0.798** \& 0.847** \& 0.798**\\ 
    \hline
    \textbf{$CoP_{ML}$} \& 0.439** \& 0.460** \& 0.800** \& 0.818** \& 0.800**\\ 
    \hline
    \textbf{$CoP_{RD}$} \& 0.405** \& 0.432** \& 0.796** \& 0.815** \& 0.796**\\ 
    \hline
    \multicolumn{6}{l}{$^{**}$ Correlation is statistically significant at the 0.001 level (2-tailed) $(p < 0.001)$.}
    \end{tabular}
    \vspace{-1.5em}
    
\end{table*}

Strong correlations were observed between the BM and FP measurements for the sway path, the sway range, and the sway velocity in all conditions, as shown in Tables \ref{tab:FPVSBMEO}, and \ref{tab:FPVSBMEC} with correlation coefficients exceeding 0.64 and $p <$ 0.001.

\subsubsection{Bland Altman Analysis}
\begin{figure}[!ht]
  \centering
  \includegraphics[width=0.5\textwidth,height=0.4\textwidth,keepaspectratio]{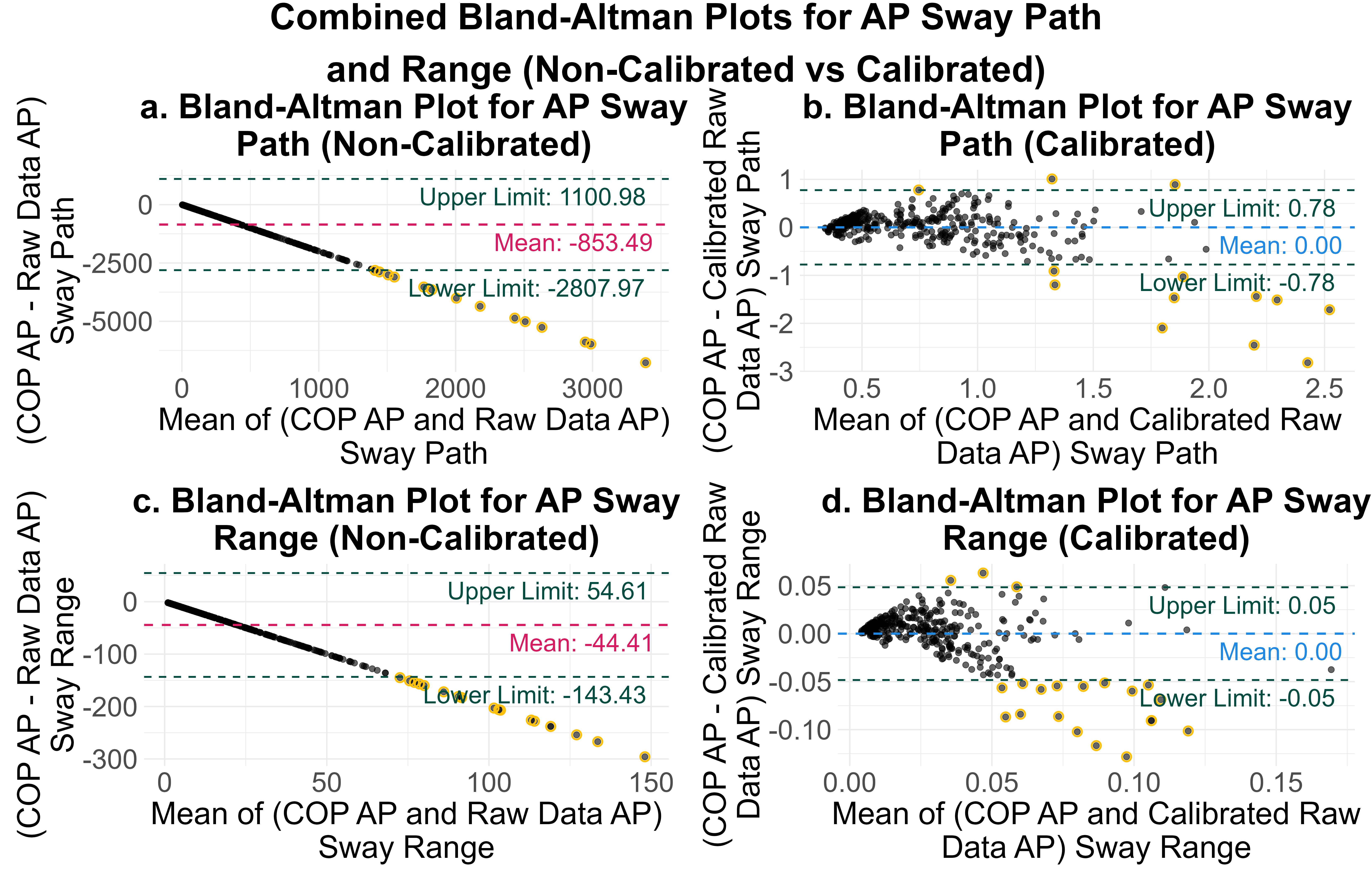}
  \caption{Bland-Altman plots for AP-direction sway metrics before and after calibration.
Figures (a) and (c) show Bland-Altman plots for the AP sway path and AP sway range before calibration. Figures (b) and (d) display the same metrics after BM raw data calibration using simple linear regression. The post-calibration plots exhibit reduced bias, with data points more symmetrically distributed around zero, indicating effective correction of systematic discrepancies. Red points represent outliers beyond the 95\% limits of agreement.}
\vspace{-1.5em}
    \label{fig:BAPlotCoPX}
\end{figure}

\begin{figure}[!ht]
  \centering
  \includegraphics[width=0.5\textwidth,height=0.4\textwidth,keepaspectratio]{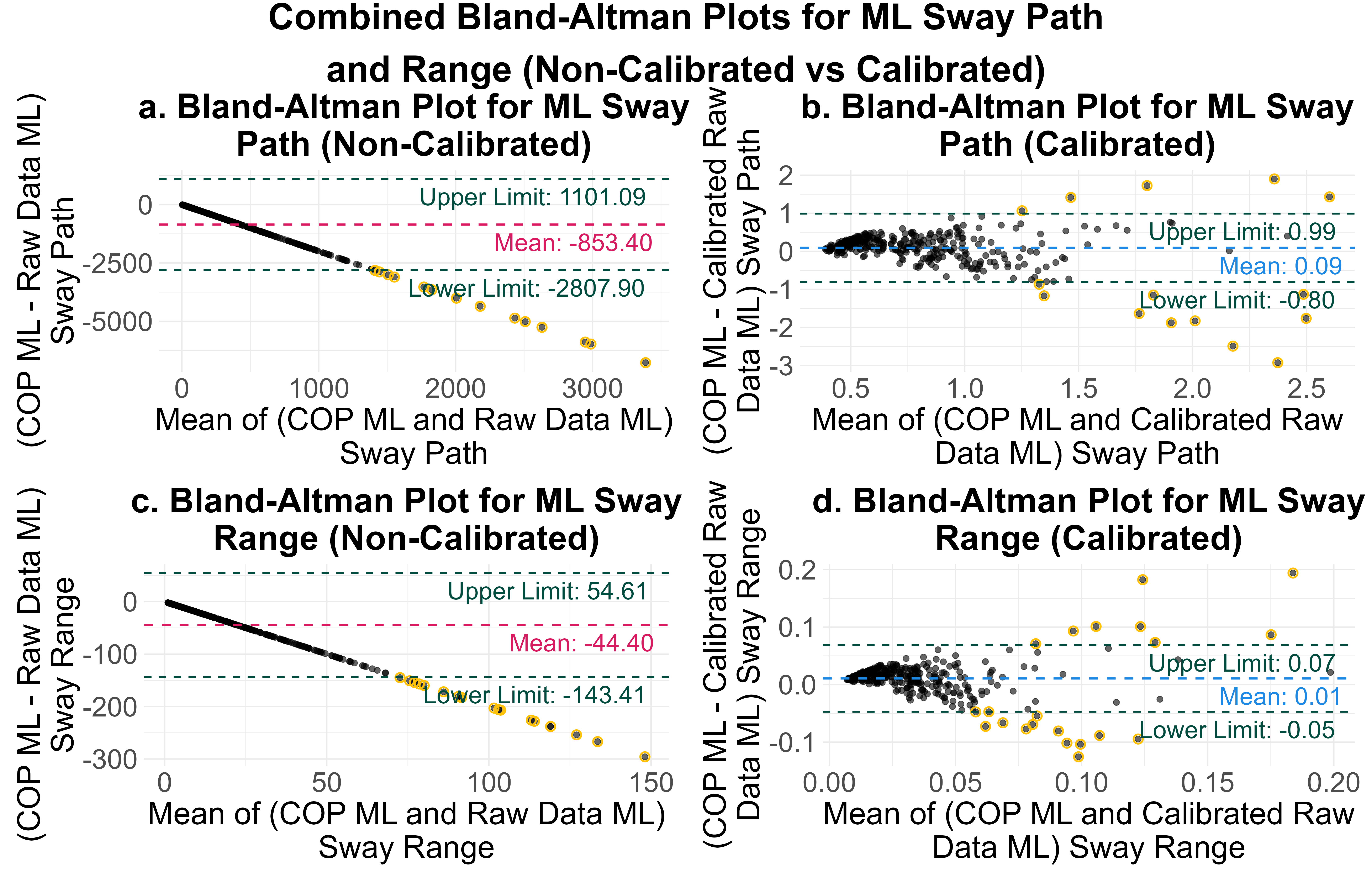}
  \caption{Bland-Altman plots for ML-direction sway metrics before and after calibration.}
Figures (a) and (c) show Bland-Altman plots for the ML sway path and ML sway range before calibration. Figures (b) and (d) display the same metrics after BM raw data calibration using simple linear regression. The post-calibration plots exhibit reduced bias, with data points more symmetrically distributed around zero, indicating effective correction of systematic discrepancies. Red points represent outliers beyond the 95\% limits of agreement.
    \vspace{-1.5em}
  \label{fig:BAPlotCoPY}
\end{figure}

Bland Altman plots were constructed to evaluate the agreement between the BM and FP measurements. Figures \ref{fig:BAPlotCoPX} and \ref{fig:BAPlotCoPY} show the plots for the CoP Sway Path and CoP Sway Range in AP and ML direction, respectively. These two parameters were chosen as they reflect the main characteristics of sway, and the Bland-Altman plots for other measures showed similar patterns. 

In the Bland-Altman plots for the $CoP_{AP}$ Sway Path, the mean difference between the BM and FP data was -853.49, with 95\% limits of agreement (LoA) ranging from 1100.98 to -2807.97. The mean difference was -44.41, with 95\% LoA from 54.61 to -143.43 for the $CoP_{AP}$ Sway Range. These results indicate that the BM consistently overestimates the sway path compared to the FP, as illustrated on the left side of Figure \ref{fig:BAPlotCoPX}. Similarly, for $CoP_{ML}$ Sway Path, the mean difference was -853.4 (95\% LoA: 1101.09 to -2807.9), and for $CoP_{ML}$ Sway Range, the mean difference was -44.4 (95\% LoA: 54.61 to -143.41), further demonstrating the BM's tendency to overestimate sway range, as shown on the left side of Figure \ref{fig:BAPlotCoPY}. A linear relationship was observed in the Bland-Altman plots, with the difference between FP and BM data increasing as the magnitude of the mean increased. Regression equations were applied to calibrate the Bland-Altman plots for both sway path and sway range, with the following equations:
\begin{itemize}
    \setlength\itemsep{0.2em}
  \item Raw Data AP Sway Path = 1928*$CoP_{AP}$ Sway Path - 624
  \item Raw Data AP Sway Range = 1580*$CoP_{AP}$ Sway Range - 1.2
  \item Raw Data ML Sway Path = 1322*$CoP_{ML}$ Sway Path - 281
  \item Raw Data ML Sway Range = 909*$CoP_{ML}$ Sway Range + 8.5
\end{itemize}
After applying calibration from simple linear regression, the accuracy of the BM measurements improved significantly for both the sway path and sway range in both the AP and ML directions. This improvement is shown on the right side of figures \ref{fig:BAPlotCoPX} and \ref{fig:BAPlotCoPY}. 

\section{Discussion and Conclusion}

This study aimed to assess the reliability and validity of the BM for measuring postural sway, utilising both robot (controlled) and human data. The results indicate that following simple linear regression adjustments, BM measures are consistent, which supports their test-retest reliability. The results from BM are also comparable to those from FP, which supports its validity against the gold standard FP.

The robot study showed that the BM has excellent test-retest reliability. Reliability was evaluated using the ICC, where values greater than 0.9 were considered excellent and values between 0.75 and 0.9 were considered good as outlined by Koo and Li \cite{koo2016guideline}. The test results showed excellent measurement consistency in the double leg stance test at 0.97 (95\% CI = 0.978 - 0.984, $p <$ 0.001) and good to excellent consistency in the single leg stance test at 0.89 (95\% CI = 0.875 - 0.911, $p <$ 0.001), except for the sway mean (ICC = 0.103 with $p = 0.103$). The slightly lower ICC for the single leg stance may be because of increased postural instability introduced in this more challenging condition due to the rig configuration. These findings from the controlled environment support that the measures captured by BM are consistent and reliable, which provides an important foundation for human testing in clinical settings.

Another key finding from the robot study was the strong relationship between mass and sway parameters obtained from the BM. The cubic model provided the best fit ($R^2>$ 0.95) for the relationship between mass and sway parameters, indicating a non-linear relationship between postural sway and mass. These results align with prior research that highlights the direct proportionality of body mass on postural stability \cite{hue2007body} but are in contrast with the findings from the study by Ku \cite{ku2012biomechanical}. This sensitivity may be important for assessing postural stability in diverse populations, such as athletes or individuals with obesity, as they may experience different sway patterns depending on their body mass. Further research is needed to understand better how body mass influences sway parameters and postural stability across different groups of people. 

The human study further validated the BM by showing strong correlations with FP measurements across multiple sway parameters, including sway path, sway range, and sway velocity. Moderate to strong ($r >$ 0.64, $p <$ 0.001) correlations were found between the BM and FP measures. Interestingly, these correlations were more pronounced in the eyes closed condition, indicating that the BM may be especially useful in more challenging balance tasks, where visual input is minimised. These results align with previous studies that show that the stability decreases in the absence of visual cues \cite{schilling2009effects}.

A linear trend was observed in the Bland Altman plot, where disagreement between the BM and the FP increased with an increase in sway magnitudes indicating proportional bias \cite{bland1986statistical, ludbrook2010confidence}. This bias suggests that the BM systematically overestimates larger sway movements and limits its use as a direct replacement for the FP. To address this bias, applying a logarithmic transformation or calibration model could help align BM outputs with those of the FP \cite{leach2014validating, sturnieks2011validity}. A simple linear regression calibration model was applied to align the BM outputs with those of the FP similar to those described in the literature by Leach et al. (2014)\cite{leach2014validating}. The calibration significantly improved the accuracy of the BM measurements for both sway path and sway range in both the AP and ML directions. 

Our results from this combined robot-human study suggest that the BM is a reliable and valid tool for measuring postural sway in healthy young adults after calibration, showing moderate to strong correlations with FP measurements and consistently capturing sway across various stance conditions. There are some limitations to this study. In the robot study, the rig configuration might not be able to fully replicate the human postural sway, which resulted in lower ICC for single leg stance. Although the robot study demonstrated excellent test-retest reliability under controlled conditions, human postural sway is inherently more variable, and additional studies are needed to evaluate reliability in real-world settings. We used controlled (robot) movement to assess the relationship between mass and postural sway, which might differ in humans and needs further exploration. We used simple linear regression equations to calibrate the BM measurements to those of the FP; more advanced calibration (e.g., machine learning-based adjustments) might be necessary for precise alignments. The human study focused primarily on healthy young adults, and further validation is needed in older populations or individuals with balance impairments to better understand the boundaries of its accuracy. Integrating AI (e.g., deep learning models trained on multi-modal balance data) could further enhance accuracy by accounting for population-specific sway patterns and automating calibration. Despite these limitations, the BM’s portability and ease of use position it as a practical tool for balance assessment in both research and clinical environments.
\vspace{-1em}




\begin{thebibliography}{99}
\bibitem{noohu2014relevance}Noohu, M., Dey, A. \& Hussain, M. Relevance of balance measurement tools and balance training for fall prevention in older adults. {\em Journal Of Clinical Gerontology And Geriatrics}. \textbf{5}, 31-35 (2014)
\bibitem{levy2018validity}Levy, S., Thralls, K. \& Kviatkovsky, S. Validity and reliability of a portable balance tracking system, BTrackS, in older adults. {\em Journal Of Geriatric Physical Therapy}. \textbf{41}, 102-107 (2018)
\bibitem{guskiewicz1996research}Guskiewicz, K. \& Perrin, D. Research and clinical applications of assessing balance. {\em Journal Of Sport Rehabilitation}. \textbf{5}, 45-63 (1996)
\bibitem{berg1992measuring}Berg, K. Measuring balance in the elderly: Development and validation of an instrument. (McGill University,1992)
\bibitem{tinetti1986performance}Tinetti, M. Performance-oriented assessment of mobility problems in elderly patients. {\em Journal Of The American Geriatrics Society}. (1986)
\bibitem{podsiadlo1991timed}Podsiadlo, D. \& Richardson, S. The timed “Up \& Go”: a test of basic functional mobility for frail elderly persons. {\em Journal Of The American Geriatrics Society}. \textbf{39}, 142-148 (1991)
\bibitem{richmond2018leveling}Richmond, S., Dames, K., Goble, D. \& Fling, B. Leveling the playing field: Evaluation of a portable instrument for quantifying balance performance. {\em Journal Of Biomechanics}. \textbf{75} pp. 102-107 (2018)
\bibitem{kimberlin2008validity}Kimberlin, C. \& Winterstein, A. Validity and reliability of measurement instruments used in research. {\em American Journal Of Health-system Pharmacy}. \textbf{65}, 2276-2284 (2008)
\bibitem{Ghahramani2022}Ghahramani, M., Hosseini, I. \& Herath, D. Performance Analysis of a Postural Balance Assessment Mat Prototype Using Inertial Sensor. {\em 2022 IEEE Sensors}. pp. 1-4 (2022)
\bibitem{leach2014validating}Leach, J., Mancini, M., Peterka, R., Hayes, T. \& Horak, F. Validating and calibrating the Nintendo Wii balance board to derive reliable center of pressure measures. {\em Sensors}. \textbf{14}, 18244-18267 (2014)
\bibitem{fritz2012effect}Fritz, C., Morris, P. \& Richler, J. Effect size estimates: current use, calculations, and interpretation.. {\em Journal Of Experimental Psychology: General}. \textbf{141}, 2 (2012)
\bibitem{prieto1996measures}Prieto, T., Myklebust, J., Hoffmann, R., Lovett, E. \& Myklebust, B. Measures of postural steadiness: differences between healthy young and elderly adults. {\em IEEE Transactions On Biomedical Engineering}. \textbf{43}, 956-966 (1996)
\bibitem{paillard2015techniques}Paillard, T. \& Noé, F. Techniques and methods for testing the postural function in healthy and pathological subjects. {\em BioMed Research International}. \textbf{2015}, 891390 (2015)
\bibitem{Mancini2012isway}Mancini, M., Salarian, A., Carlson-Kuhta, P., Zampieri, C., King, L., Chiari, L. \& Horak, F. ISway: a sensitive, valid and reliable measure of postural control. {\em Journal Of Neuroengineering And Rehabilitation}. \textbf{9} pp. 1-8 (2012)
\bibitem{muir2013dynamic}Muir, J., Kiel, D., Hannan, M., Magaziner, J. \& Rubin, C. Dynamic parameters of balance which correlate to elderly persons with a history of falls. {\em Plos One}. \textbf{8}, e70566 (2013)
\bibitem{fernie1982relationship}Fernie, G., Gryfe, C., Holliday, P. \& Llewellyn, A. The relationship of postural sway in standing to the incidence of falls in geriatric subjects. {\em Age And Ageing}. \textbf{11}, 11-16 (1982)
\bibitem{sturnieks2011validity}Sturnieks, D., Arnold, R. \& Lord, S. Validity and reliability of the Swaymeter device for measuring postural sway. {\em BMC Geriatrics}. \textbf{11} pp. 1-7 (2011)
\bibitem{schober2018correlation}Schober, P., Boer, C. \& Schwarte, L. Correlation coefficients: appropriate use and interpretation. {\em Anesthesia \& Analgesia}. \textbf{126}, 1763-1768 (2018)
\bibitem{koo2016guideline}Koo, T. \& Li, M. A guideline of selecting and reporting intraclass correlation coefficients for reliability research. {\em Journal Of Chiropractic Medicine}. \textbf{15}, 155-163 (2016)
\bibitem{hue2007body}Hue, O., Simoneau, M., Marcotte, J., Berrigan, F., Doré, J., Marceau, P., Marceau, S., Tremblay, A. \& Teasdale, N. Body weight is a strong predictor of postural stability. {\em Gait \& Posture}. \textbf{26}, 32-38 (2007)
\bibitem{ku2012biomechanical}Ku, P., Osman, N., Yusof, A. \& Abas, W. Biomechanical evaluation of the relationship between postural control and body mass index. {\em Journal Of Biomechanics}. \textbf{45}, 1638-1642 (2012)
\bibitem{schilling2009effects}Schilling, B., Falvo, M., Karlage, R., Weiss, L., Lohnes, C. \& Chiu, L. Effects of unstable surface training on measures of balance in older adults. {\em The Journal Of Strength \& Conditioning Research}. \textbf{23}, 1211-1216 (2009)
\bibitem{bland1986statistical}Bland, J. \& Altman, D. Statistical methods for assessing agreement between two methods of clinical measurement. {\em The Lancet}. \textbf{327}, 307-310 (1986)
\bibitem{ludbrook2010confidence}Ludbrook, J. Confidence in Altman–Bland plots: a critical review of the method of differences. {\em Clinical And Experimental Pharmacology And Physiology}. \textbf{37}, 143-149 (2010)
\end{thebibliography}

\end{document}